\documentclass[preprintnumbers,amsmath,11pt,amssymb,floatfix,superscriptaddress,nofootinbib]{article}

\def\mytitle#1{\setcounter{equation}{0}
\setcounter{footnote}{0}
\begin{flushleft}\Large\textbf{#1}\end{flushleft}
\vspace{0.25cm}}
\def\myname#1{\leftline{{\large #1}}\vspace{-0.13cm}}
\def\myplace#1#2{\small\begin{flushleft}\textit{#1}\\
\texttt{#2}\end{flushleft}}

\usepackage{graphicx}
\begin{document}

\mytitle{Possibility of cyclic Turnarounds In Brane-world
Scenario: Phantom Energy Accretion onto Black Holes and its
consequences.}

\vskip0.2cm \myname{ Prabir
Rudra~\footnote{prudra.math@gmail.com}}

\myplace{Department of Mathematics, Bengal Engineering and Science
University, Shibpur, Howrah-711 103, India.} {}

\begin{abstract}
A universe described by braneworlds is studied in a cyclic
scenario. As expected such an oscillating universe will undergo
turnarounds, whenever the phantom energy density reaches a
critical value from either side. It is found that a universe
described by RSII brane model will readily undergo oscillations
if, either the brane tension, $\lambda$ or the bulk cosmological
constant, $\Lambda_{4}$ is negative. The DGP brane model does not
readily undergo cyclic turnarounds. Hence for this model a
modified equation is proposed to incorporate the cyclic nature. It
is found that there is always a remanent mass of a black hole at
the verge of a turnaround. Hence contrary to known results in
literature, it is found that the destruction of black holes at the
turnaround is completely out of question. Finally to alleviate, if
not solve, the problem posed by the black holes, it is argued that
the remanent masses of the black holes do not act as a serious
defect of the model because of Hawking evaporation.

\end{abstract}

\section{Introduction}

\vspace{3mm}

Cyclic universe has always been a burning topic in the field of
theoretical cosmology, since it is expected to avoid the initial
singularity by providing an infinitely oscillating universe.
However cyclic universe confront a serious problem of black
holes(BHs). If the BHs formed during the expanding phase survives
into the next cycle they will grow even larger from one cycle to
the next and act as a serious defect in an otherwise nearly
uniform universe. With the passage of time the BHs will occupy the
entire horizon and then the cyclic models will break away. In this
paper we investigate the possibility of an oscillating universe in
two of the well known models of brane-world gravity, namely, RSII
brane and DGP brane models.

\vspace{3mm}

Randall and Sundrum \cite{Randall1, Randall2} proposed a
bulk-brane model to explain the higher dimensional theory,
popularly known as RS II brane model. According to this model we
live in a four dimensional world (called 3-brane, a domain wall)
which is embedded in a 5D space time (bulk). All matter fields are
confined in the brane whereas gravity can only propagate in the
bulk. The consistency of this brane model with the expanding
universe has given popularity to this model of late in the field
of cosmology.

\vspace{3mm}

A simple and effective model of brane-gravity is the
Dvali-Gabadadze-Porrati (DGP) braneworld model \cite{Dvali1} which
models our 4-dimensional world as a FRW brane embedded in a
5-dimensional Minkowski bulk. It explains the origin of dark
energy(DE) as the gravity on the brane leaking to the bulk at
large scale. On the 4-dimensional brane the action of gravity is
proportional to $M_p^2$ whereas in the bulk it is proportional to
the corresponding quantity in 5-dimensions. The model is then
characterized by a cross over length scale $
r_c=\frac{M_p^2}{2M_5^2} $ such that gravity is 4-dimensional
theory at scales $a<<r_c$ where matter behaves as pressureless
dust, but gravity leaks out into the bulk at scales $a>>r_c$ and
matter approaches the behaviour of a cosmological constant.
Moreover it has been shown that the standard Friedmann cosmology
can be firmly embedded in DGP brane.

\vspace{3mm}

To explain the latest cosmic acceleration one usually assumes the
existence of dark energy (DE) with a negative pressure. In general
one can assume a perfect fluid with state equation $p=\omega\rho$,
with $\omega<-\frac{1}{3}$, in order to realize the cosmic
acceleration. Most models of DE in the present Universe predict
that its effective equation of state satisfies the null energy
condition (NEC), i.e.,  $\omega_{eff} = p_{DE}/\rho_{DE}\geq-1$,
where $\rho_{DE}$ and $p_{DE}$ are the effective DE density and
pressure, respectively. However, the observations do not rule out
that DE is phantom, i.e., it violates NEC. Observations from WMAP
indicates the value $\omega=-1.10$ \cite{Kamatsu1}, which means
that our universe would be dominated by 'phantom energy'
($\omega<-1$). It has been shown in \cite{gr} that phantom dark
energy can be successfully accomodated within framework of General
Relativity (GR).

\vspace{3mm}

In the context of BHs and phantom energy accretion on BH, it
should be mentioned that Babichev et al \cite{Babichev1} has shown
that BH mass decrease with phantom energy accretion on it. Jamil
et al \cite{Jamil1} studied charged BHs in phantom cosmology.
Jamil in \cite{Jamil2} has shown the evolution of a Schwarzschild
Black Hole in Phantom-like Chaplygin gas Cosmologies. Primordial
BHs in phantom cosmology and accretion of phantom DE on BTZ BHs
were also studied by Jamil et al in \cite{Jamil3, Jamil4}. Nayak
in \cite{Nayak1} investigated the effect of Vacuum Energy on the
evolution of primordial BHs in Einstein Gravity. Paolis in
\cite{Paolis1} studied BHs in bulk viscous cosmology. In the
context of cyclic cosmology, it should be mentioned that Saridakis
in \cite{Saridakis1} studied cyclic Universes from general
collisionless Braneworld models. Cai et al in \cite{Cai1}
investigated cyclic extension of the non-singular cosmology in a
model of non-relativistic gravity. Cai et al in \cite{Cai2}
investigated cyclic and singularity-free evolutions in a universe
governed by Lagrange-multiplier modified gravity. Moreover Cai et
al in \cite{Cai3} showed that gravity described by an arbitrary
function of the torsion scalar, can provide a mechanism for
realizing bouncing cosmologies, thereby avoiding the Big Bang
singularity. Non-singular cyclic cosmology without phantom menace
was also studied by Cai et al in \cite{Cai4}.

\vspace{3mm}

We intend to study the effects  and consequences of phantom energy
accretion onto BHs in a cyclic scenario of the universe described
by DGP and RSII branes. Our motivation is to find out if there is
any remanent mass of BH when it undergoes a turnaround in a cyclic
scenario. As mentioned earlier Babichev et al \cite{Babichev1} has
shown that BH mass decrease with phantom energy accretion on it.
Hence the BH will disappear before the turnaround in an
oscillating universe. But Sun \cite{Sun1} provided a mechanism
which showed that in an universe described by modified Friedmann
cosmology the destruction of BHs is totally out of question, as
there is always a remanent mass of a BH facing a turnaround. In
this paper our motivation is to testify the above fact for
brane-world cosmology and find out the fate of a BH undergoing
phantom energy accretion in an oscillating universe.

\vspace{3mm}

The paper is organised as follows:  In section 2 we discuss the
mechanism of cyclic universe in RSII brane model. Section 3 deals
with an identical mechanism for DGP brane model. In section 4, we
present an argument regarding Hawking evaporation of remanent BHs.
Finally the paper ends with some concluding remarks in section 5.

\vspace{3mm}

\section{Cyclic Universe in RSII Brane Model}

 The novel feature of the RS models compared to
previous higher-dimensional models is that the observable 3
dimensions are protected from the large extra dimension (at low
energies) by curvature rather than straightforward
compactification. In RS II model the effective equations of motion
on the 3-brane embedded in 5D bulk having $Z_{2}$-symmetry are
given by \cite{Maartens1, Maartens2, Randall1, Shiromizu1, Maeda1,
Sasaki1}
\begin{equation}\label{cyclic5.1}
^{(4)}G_{\mu\nu}=-\Lambda_{4}q_{\mu\nu}+\kappa^{2}_{4}\tau_{\mu\nu}+\kappa^{4}_{5}\Pi_{\mu\nu}-E_{\mu\nu}
\end{equation}
where

\begin{equation}\label{cyclic5.2}
\kappa^{2}_{4}=\frac{1}{6}~\lambda\kappa^{4}_{5}~,
\end{equation}
\begin{equation}\label{cyclic5.3}
\Lambda_{4}=\frac{1}{2}~\kappa^{2}_{5}\left(\Lambda_{5}+\frac{1}{6}~\kappa^{2}_{5}\lambda^{2}\right)
\end{equation}
and
\begin{equation}\label{cyclic5.4}
\Pi_{\mu\nu}=-\frac{1}{4}~\tau_{\mu\alpha}\tau^{\alpha}_{\nu}+\frac{1}{12}~\tau\tau_{\mu\nu}+\frac{1}{8}~
q_{\mu\nu}\tau_{\alpha\beta}\tau^{\alpha\beta}-\frac{1}{24}~q_{\mu\nu}\tau^{2}
\end{equation}
and $E_{\mu\nu}$ is the electric part of the 5D Weyl tensor. Here
$\kappa_{5},~\Lambda_{5}, \lambda, ~\tau_{\mu\nu}$ and
$\Lambda_{4}$ are respectively the 5D gravitational coupling
constant, 5D cosmological constant, the brane tension (vacuum
energy), brane energy-momentum tensor and effective 4D
cosmological constant. The explicit form of the above modified
Einstein equations in flat universe for RSII brane are

\begin{equation}\label{cyclic5.5}
3H^{2}=\Lambda_{4}+\kappa^{2}_{4}\rho+\frac{\kappa^{2}_{4}}{2\lambda}~\rho^{2}+\frac{6}{\lambda
\kappa^{2}_{4}}\cal{U}
\end{equation}
and
\begin{equation}\label{cyclic5.6}
2\dot{H}+3H^{2}=\Lambda_{4}-\kappa^{2}_{4}p-\frac{\kappa^{2}_{4}}{2\lambda}~\rho
p-\frac{\kappa^{2}_{4}}{2\lambda}~\rho^{2}-\frac{2}{\lambda
\kappa^{2}_{4}}\cal{U}
\end{equation}
The dark radiation $\cal{U}$ obeys

\begin{equation}\label{cyclic5.7}
\dot{\cal U}+4H{\cal U}=0
\end{equation}
where $\rho$ and $p$ are the total energy density and pressure
respectively.

\subsection{Phantom Energy Accretion}

We consider an homogeneous and isotropic universe filled with DE
fluid, with DE density $\rho$ and pressure $p$. For an asymptotic
observer the black hole mass, $M$ changes at the rate of
\cite{Babichev1}
\begin{equation}\label{cyclic5.8}
\dot{M}=4\pi AM^{2}\left(\rho+p\right)
\end{equation}
Here the overdot denotes the derivative with respect to cosmic
time. Moreover it has been considered that $G=c=1$. We consider an
universe dominated by DE. The Friedmann equation for the expanding
universe is given by
\begin{equation}\label{cyclic5.9}
H^{2}=\frac{8\pi}{3}\rho
\end{equation}
Where $H\equiv\frac{\dot{a}}{a}$ is the Hubble parameter. The
conservation equation for DE is given by
\begin{equation}\label{cyclic5.10}
\dot{\rho}+3H\left(\rho+p\right)=0
\end{equation}
Phantom DE has the equation of state ~$w=\frac{p}{\rho}<-1$. We
get that $\rho\propto a^{-3(1+w)}$, which shows that $\rho$
increases with the expansion of the universe.

In this section we will investigate the dynamics of cyclic
universe in RSII brane model. The modified Friedmann equation for
RSII brane model is given by equation (\ref{cyclic5.5}). Putting $
{\cal{U}}=0 $,  the equation becomes,
\begin{equation}\label{cyclic5.11}
H^{2}=\frac{\kappa_{4}^{2}}{3}\rho\left(1+\frac{\rho}{2\lambda}\right)+\frac{\Lambda_{4}}{3}
\end{equation}
In the expanding phase of the universe, the phantom energy
density, $\rho$ increases. From the above equation we see that at
the turnaround the phantom energy density is given by,
\begin{equation}\label{cyclic5.12}
\rho_{c}=-\lambda\pm \sqrt{\lambda^{2}-\alpha}
\end{equation}
where $\alpha=\frac{2\lambda\Lambda_{4}}{\kappa_{4}^{2}}$.

From the above value we see that when $\lambda>0$, then
$\Lambda_{4}<0$, and when $\lambda<0$, then $\Lambda_{4}>0$.
$\lambda>\frac{2\Lambda_{4}}{\kappa_{4}^{2}}$. This shows that
either the brane tension or the bulk cosmological constant has to
be negative, so that the universe undergoes a bounce, and a
possibility for the cyclic scenario is evident. Here $\rho_{c}$
stands for critical density. After the turnaround the universe
begins to contract. The reason behind this being, that in the
contracting phase the non-phantom components of the universe
increase and begins to dominate the evolution. In this phase,
again when the dominant energy density reaches the critical value
given by equation (\ref{cyclic5.12}), a bounce occurs. Thus we get
an oscillating scenario. We intend to study the variation of BH
mass with phantom energy accretion around it, in this oscillating
cosmological model.

\subsection{Scenario Before Turnaround}
Using equations (\ref{cyclic5.8}) and (\ref{cyclic5.10}) we get
\begin{equation}\label{cyclic5.13}
\frac{dM}{M^2}=-\frac{4\pi A}{3H}d\rho
\end{equation}
Before turnaround, we have
\begin{equation}\label{cyclic5.14}
H=\frac{\kappa_{4}}{\sqrt{3}}\sqrt{\rho\left(1+\frac{\rho}{2\lambda}\right)+\frac{\Lambda_{4}}{\kappa_{4}^{2}}}
\end{equation}
Now substituting the above value of $H$ in equation
(\ref{cyclic5.13}) we get,
\begin{equation}\label{cyclic5.15}
\frac{dM}{M^2}=-\frac{D}{\sqrt{\rho\left(1+\frac{\rho}{2\lambda}\right)+\frac{\Lambda_{4}}{\kappa_{4}^{2}}}}d\rho
\end{equation}
where $D=\frac{4\pi A}{\sqrt{3}\kappa_{4}}$ and $\rho_{c}$ is
given by equation (\ref{cyclic5.12}). Now integrating equation
(\ref{cyclic5.15}) we get,
\begin{equation}\label{cyclic5.16}
M=\frac{M_{i}}{1+DM_{i}\sqrt{2\lambda}\log{\left(\frac{\lambda+\rho+\sqrt{\alpha+2\lambda\rho+\rho^{2}}}{\lambda+\rho_{i}+\sqrt{\alpha+2\lambda\rho_{i}+\rho_{i}^{2}}}\right)}}
\end{equation}
Here $\rho_{i}$ and $M_{i}$ denotes respectively the phantom
energy density and the black hole mass at the moment when the
phantom energy density just begins to dominate the evolution of
the universe. In general, $\rho_{i}\ll\rho_{c}$ and
$\rho_{i}\leq\rho\leq\rho_{c}$. So using equation
(\ref{cyclic5.12}) we obtain
$\frac{\rho_{i}}{\rho_{c}}\rightarrow0$. Hence from equation
(\ref{cyclic5.16}), we obtain
\begin{equation}\label{cyclic5.17}
M\simeq\frac{M_{i}}{1+DM_{i}\sqrt{2\lambda}\log{\left(\frac{\lambda+\rho+\sqrt{\alpha+2\lambda\rho+\rho^{2}}}{\lambda}\right)}}
\end{equation}
At the turnaround, $\rho=\rho_{c}$ and hence the black hole mass
at the turnaround becomes,
\begin{equation}\label{cyclic5.18}
M_{c}\simeq\frac{M_{i}}{1+DM_{i}\sqrt{2\lambda}\log{\left(\frac{\lambda+\rho_{c}+\sqrt{\alpha+2\lambda\rho_{c}+\rho_{c}^{2}}}{\lambda}\right)}}
\end{equation}
This shows that there is a remnant mass of the BH when the
turnaround occurs. Hence this result is different from the result
obtained by Zhang \cite{Zhang1}. It is quite clear from the above
equations that initially, through phantom energy accretion, the BH
mass decreases, until it reaches the minimum value $M_{c}$ at the
turnaround in the expanding phase. For $M_{i}\gg
M_{p}=G^{-\frac{1}{2}}$, $M_{c}$ becomes independent of $M_{i}$
\begin{equation}\label{cyclic5.19}
M_{c}\simeq\frac{1}{D\sqrt{2\lambda}\log{\left(\frac{\lambda+\rho_{c}+\sqrt{\alpha+2\lambda\rho_{c}+\rho_{c}^{2}}}{\lambda}\right)}}
\end{equation}

\subsection{Scenario After Turnaround}
After the turnaround as expected the the universe will contract
and consequently the phantom energy density $\rho$ starts
decreasing. During the contraction it is obvious that the time
derivative of the scale factor will become negative, and as a
result $H$ becomes negative. So, here we take the value of $H$ as,
\begin{equation}\label{cyclic5.20}
H=-\frac{\kappa_{4}}{\sqrt{3}}\sqrt{\rho\left(1+\frac{\rho}{2\lambda}\right)+\frac{\Lambda_{4}}{\kappa_{4}^{2}}}
\end{equation}
Substituting the above value of $H$ in equation (\ref{cyclic5.13})
we get,
\begin{equation}\label{cyclic5.21}
\frac{dM}{M^2}=\frac{D}{\sqrt{\rho\left(1+\frac{\rho}{2\lambda}\right)+\frac{\Lambda_{4}}{\kappa_{4}^{2}}}}d\rho
\end{equation}
Now integrating the above equation we get,
\begin{equation}\label{cyclic5.22}
M=\frac{M_{c}}{1+DM_{c}\sqrt{2\lambda}\log{\left(\frac{\lambda+\rho_{c}+\sqrt{\alpha+2\lambda\rho_{c}+\rho_{c}^{2}}}{\lambda+\rho+\sqrt{\alpha+2\lambda\rho+\rho^{2}}}\right)}}
\end{equation}
where $\rho\leq\rho_{c}$. The above equation shows that as the
universe contracts, the BH mass continue to decrease. When
$\rho\ll\rho_{c}$, then the BH mass is,
\begin{equation}\label{cyclic5.23}
M_{f}\simeq\frac{M_{c}}{1+DM_{c}\sqrt{2\lambda}\log{\left(\frac{\lambda+\rho_{c}+\sqrt{\alpha+2\lambda\rho_{c}+\rho_{c}^{2}}}{\lambda}\right)}}
\end{equation}
For $M_{i}\gg M_{p}$, using equation (\ref{cyclic5.19}) we find
that the final mass of BHs is
\begin{equation}\label{cyclic5.24}
M_{f}\simeq \frac{M_{c}}{2}\simeq
\frac{1}{2D\sqrt{2\lambda}\log{\left(\frac{\lambda+\rho_{c}+\sqrt{\alpha+2\lambda\rho_{c}+\rho_{c}^{2}}}{\lambda}\right)}}
\end{equation}
Hence $M_{f}$ is independent of $M_{i}$.

\begin{figure}

\includegraphics[height=2in]{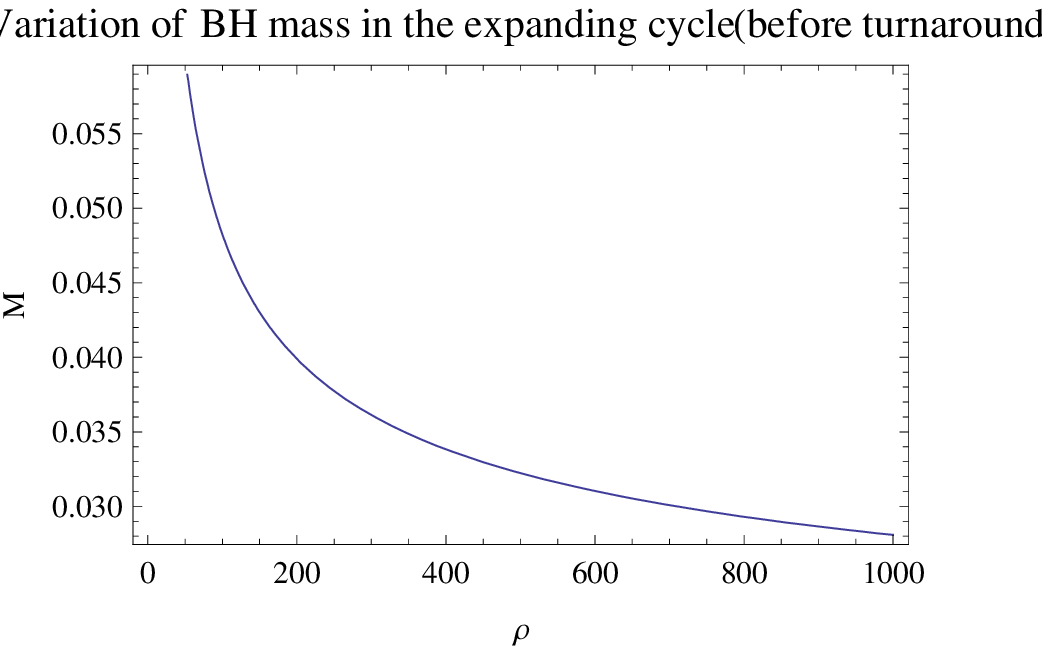}~~~~~~~~\includegraphics[height=2in]{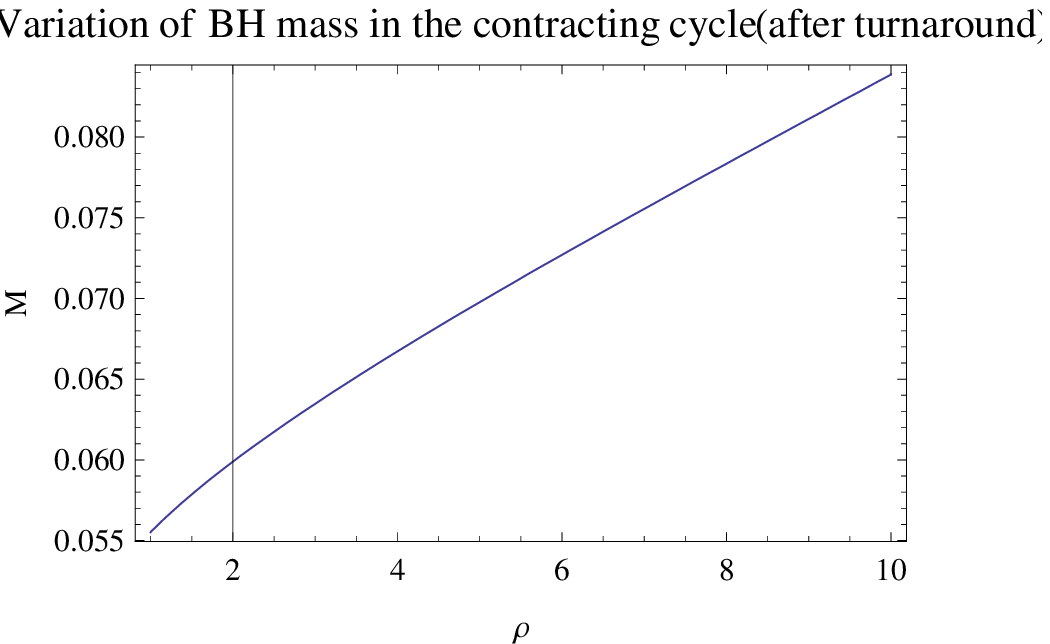}~~~~~\\
\vspace{2mm}
~~~~~~~~~~~~~~~~~~~~~~~~Fig. 1~~~~~~~~~~~~~~~~~~~~~~~~~~~~~~~~~~~~~~~~~~~~~~~~~~~~~~~~~~~~~~Fig. 2~~~~~\\
\vspace{2mm}

Fig 1 : The mass of the black hole is plotted against the
increasing density of phantom dark energy before turnaround. Other
parameters are fixed at $\alpha=-5, \lambda=10, D=1.5, M_{i}=10000$, $\rho_{i}=0.0001$ \\

Fig 2 : The mass of the black hole is plotted against the
decreasing density of phantom dark energy after turnaround. Other
parameters are fixed at $\alpha=-5, \lambda=10, D=1.5, M_{c}=100, \rho_{c}=100$\\

\end{figure}

\noindent

In figure 1, plot is generated for the mass of the BH against the
density of phantom DE, in the expanding cycle before the
turnaround. We know that in the expanding phase, the phantom
energy components dominates over the non-phantom counterparts. As
a result, phantom energy begins accreting on the BHs in this
cycle. From the figure, it is seen that with the increase in the
phantom energy density before turnaround, the BH mass gradually
decreases, due to more and more accreting phenomenon going on the
BH. This result is in complete accordance with Babichev et al
\cite{Babichev1}.

\noindent

In figure 2, the BH mass is plotted against the phantom energy
density, in the contracting cycle after the turnaround. It is
clearly evident from the figure that as the non-phantom components
begin to dominate in the contracting phase, the BH mass cease to
diminish. This is quite an expected result.

\section{Cyclic Universe in DGP Brane Model}
While flat, homogeneous and isotropic brane is being considered,
the Friedmann equation in DGP brane model is modified to the
equation
\begin{equation}\label{cyclic5.25}
H^2=\left(\sqrt{\frac{\rho}{3}+\frac{1}{4r_{c}^{2}}}+\epsilon
\frac{1}{2r_c}\right)^2
\end{equation}
where $H=\frac{\dot a}{a}$ is the Hubble parameter, $\rho$ is the
total cosmic fluid energy density and $r_c=\frac{M_p^2}{2M_5^2}$
is the cross-over scale which determines the transition from 4D to
5D behaviour and $\epsilon=\pm 1 $ (choosing $M_{p}^{2}=8\pi
G=1$). For $\epsilon=+1$, we have standard DGP$(+)$ model which is
self accelerating model without any form of DE, and effective $w$
is always non-phantom. However for $\epsilon=-1$, we have DGP$(-)$
model which does not self accelerate but requires DE on the brane.
It experiences 5D gravitational modifications to its dynamics
which effectively screen DE.

Like the original Friedmann equation, and contrary to RSII brane
equation, the DGP brane equation does not readily support an
oscillating universe undergoing turnarounds. This fact is quite
obvious from the equation (\ref{cyclic5.25}). We can see that
there is no possibility of turnaround. Therefore we will propose a
modified DGP brane equation, that can support a cyclic universe.
Sun et al in \cite{Sun2} proposed a method, by which they were
able to modify the Friedmann equation into a form that avoids the
Big Rip singularity and gives a bouncing cosmological model. Their
investigations led to the fact that the character of physics
changes remarkably near the Planck scale. We know that a de Sitter
universe with a cosmological constant, $\Lambda$ is similar to a
BH. Also it has a temperature, $T\sim H$. Hence by conjecturing
physics at the Planck scale, they actually modified the definition
of Hawking temperature, and subsequently obtained the modified
Friedmann equation. Sun in \cite{Sun1} used this modified equation
successfully to study the phantom energy accretion on BHs. Hence,
taking a leaf out of their book, we proceed to modify the DGP
brane equation as follows: The proposed modified DGP brane
equation is
\begin{equation}\label{cyclic5.26}
H^2=\left(\sqrt{\frac{\rho}{3}+\frac{1}{4r_{c}^{2}}}+\epsilon
\frac{1}{2r_c}\right)^2\left(1-\frac{\rho}{\rho_{c}'}\right)
\end{equation}
From the above equation, we see that the turnaround occurs for
$\rho=\rho_{c}'$.

\subsection{Scenario Before Turnaround}
The expression for $H$ for the expanding phase is given by,
\begin{equation}\label{cyclic5.27}
H=\left(\sqrt{\frac{\rho}{3}+\frac{1}{4r_{c}^{2}}}+\epsilon
\frac{1}{2r_c}\right)\sqrt{\left(1-\frac{\rho}{\rho_{c}'}\right)}
\end{equation}

Using equation (\ref{cyclic5.13}) and (\ref{cyclic5.27}) we get,
\begin{equation}\label{cyclic5.28}
\frac{dM}{M^{2}}=-\frac{D'}{\left(\sqrt{\frac{\rho}{3}+\frac{1}{4r_{c}^{2}}}+\frac{\epsilon}{2r_{c}}\right)\sqrt{1-\frac{\rho}{\rho_{c}'}}}
\end{equation}
where $D'=\frac{4\pi A}{3}$. Now integrating the above equation we
get,

$$\frac{1}{M}=\frac{1}{M_{i}}+D'\left[\sqrt{3\rho_{c}'}\left(\arctan\left(\frac{3A-\rho_{c}'+2\rho}{2\sqrt{\left(\rho_{c}'-\rho\right)\left(3A-\rho\right)}}\right)-\arctan\left(\frac{3A-\rho_{c}'+2\rho_{i}}{2\sqrt{\left(\rho_{c}'-\rho_{i}\right)\left(3A-\rho_{i}\right)}}\right)\right)\right.$$
$$\left.+\frac{3B\sqrt{\rho_{c}'}}{\sqrt{3B^{2}-3A-\rho_{c}'}}\left(\arctan\left(\frac{9A^{2}+3B^{2}\left(\rho_{c}'-2\rho\right)+\rho\rho_{c}'+3A\left(\rho+\rho_{c}'-3B^{2}\right)}{2B\sqrt{3\rho_{c}'\left(3A+\rho\right)\left(3B^{2}-3A-\rho_{c}'\right)\left(1-\frac{\rho}{\rho_{c}'}\right)}}\right)\right.\right.$$
$$\left.\left.-\arctan\left(\frac{9A^{2}+3B^{2}\left(\rho_{c}'-2\rho_{i}\right)+\rho_{i}\rho_{c}'+3A\left(\rho_{i}+\rho_{c}'-3B^{2}\right)}{2B\sqrt{3\rho_{c}'\left(3A+\rho_{i}\right)\left(3B^{2}-3A-\rho_{c}'\right)\left(1-\frac{\rho_{i}}{\rho_{c}'}\right)}}\right)\right)\right.$$
\begin{equation}\label{cyclic5.29}
\left.+\frac{6B\sqrt{\rho_{c}'}}{\sqrt{3A-3B^{2}+\rho_{c}'}}\left(\arctan\left(\frac{\sqrt{\rho_{c}'\left(1-\frac{\rho}{\rho_{c}'}\right)}}{\sqrt{3A-3B^{2}+\rho_{c}'}}\right)-\arctan\left(\frac{\sqrt{\rho_{c}'\left(1-\frac{\rho_{i}}{\rho_{c}'}\right)}}{\sqrt{3A-3B^{2}+\rho_{c}'}}\right)\right)\right]
\end{equation}
Where $A=\frac{1}{4r_{c}^{2}}$, and ~~$B=\frac{\epsilon}{2r_{c}}$

Now as in the case of RSII brane, $\rho_{i}\ll\rho_{c}'$, which
implies that $\frac{\rho_{i}}{\rho_{c}'}\rightarrow 0$. Hence from
the above equation we obtain,

$$\frac{1}{M}\simeq\frac{1}{M_{i}}+D'\left[\sqrt{3\rho_{c}'}\left(\arctan\left(\frac{3A-\rho_{c}'+2\rho}{2\sqrt{\left(\rho_{c}'-\rho\right)\left(3A-\rho\right)}}\right)-\arctan\left(\frac{3A-\rho_{c}'}{2\sqrt{3A\rho_{c}'}}\right)\right)\right.$$
$$\left.+\frac{3B\sqrt{\rho_{c}'}}{\sqrt{3B^{2}-3A-\rho_{c}'}}\left(\arctan\left(\frac{9A^{2}+3B^{2}\left(\rho_{c}'-2\rho\right)+\rho\rho_{c}'+3A\left(\rho+\rho_{c}'-3B^{2}\right)}{2B\sqrt{3\rho_{c}'\left(3A+\rho\right)\left(3B^{2}-3A-\rho_{c}'\right)\left(1-\frac{\rho}{\rho_{c}'}\right)}}\right)\right.\right.$$
$$\left.\left.-\arctan\left(\frac{9A^{2}+3B^{2}\rho_{c}'+3A\left(\rho_{c}'-3B^{2}\right)}{2B\sqrt{9A\rho_{c}'\left(3B^{2}-3A-\rho_{c}'\right)}}\right)\right)\right.$$
\begin{equation}\label{cyclic5.30}
\left.+\frac{6B\sqrt{\rho_{c}'}}{\sqrt{3A-3B^{2}+\rho_{c}'}}\left(\arctan\left(\frac{\sqrt{\rho_{c}'\left(1-\frac{\rho}{\rho_{c}'}\right)}}{\sqrt{3A-3B^{2}+\rho_{c}'}}\right)-\arctan\left(\frac{\sqrt{\rho_{c}'}}{\sqrt{3A-3B^{2}+\rho_{c}'}}\right)\right)\right]
\end{equation}
At the turnaround $\rho=\rho_{c}'$. Hence we get,
$$\frac{1}{M_{c}}\simeq\frac{1}{M_{i}}+D'\left[\frac{\pi}{2}\sqrt{3\rho_{c}'}-\arctan\left(\frac{3A-\rho_{c}'}{2\sqrt{3A\rho_{c}'}}\right)\right.$$
$$\left.+\frac{3B\sqrt{\rho_{c}'}}{\sqrt{3B^{2}-3A-\rho_{c}'}}\left(\frac{\pi}{2}-\arctan\left(\frac{9A^{2}+3B^{2}\rho_{c}'+3A\left(\rho_{c}'-3B^{2}\right)}{2B\sqrt{9A\rho_{c}'\left(3B^{2}-3A-\rho_{c}'\right)}}\right)\right)\right.$$
\begin{equation}\label{cyclic5.31}
\left.-\frac{6B\sqrt{\rho_{c}'}}{\sqrt{3A-3B^{2}+\rho_{c}'}}\arctan\left(\frac{\sqrt{\rho_{c}'}}{\sqrt{3A-3B^{2}+\rho_{c}'}}\right)\right]
\end{equation}
The above equation gives the remanent mass of the BH at the
turnaround. Hence we see that the mass of BH gradually decrease
during the expanding phase of cyclic universe, due to phantom
energy accretion and finally becomes minimum at the turnaround,
when it is called the critical mass, given by
equation(\ref{cyclic5.30}). For $M_{i}\gg M_{p}$, $M_{c}$ becomes
independent of $M_{i}$ and is given by the following relation,
$$\frac{1}{M_{c}}\simeq D'\left[\frac{\pi}{2}\sqrt{3\rho_{c}'}-\arctan\left(\frac{3A-\rho_{c}'}{2\sqrt{3A\rho_{c}'}}\right)\right.$$
$$\left.+\frac{3B\sqrt{\rho_{c}'}}{\sqrt{3B^{2}-3A-\rho_{c}'}}\left(\frac{\pi}{2}-\arctan\left(\frac{9A^{2}+3B^{2}\rho_{c}'+3A\left(\rho_{c}'-3B^{2}\right)}{2B\sqrt{9A\rho_{c}'\left(3B^{2}-3A-\rho_{c}'\right)}}\right)\right)\right.$$
\begin{equation}\label{cyclic5.32}
\left.-\frac{6B\sqrt{\rho_{c}'}}{\sqrt{3A-3B^{2}+\rho_{c}'}}\arctan\left(\frac{\sqrt{\rho_{c}'}}{\sqrt{3A-3B^{2}+\rho_{c}'}}\right)\right]
\end{equation}

\subsection{Scenario After Turnaround}
Just like the RSII model, in DGP model as well, the universe
starts to contract after turnaround, and the non-phantom
components starts to dominate the evolution. The expression for
$H$ is given by,
\begin{equation}\label{cyclic5.33}
H=-\left(\sqrt{\frac{\rho}{3}+\frac{1}{4r_{c}^{2}}}+\epsilon
\frac{1}{2r_c}\right)\sqrt{\left(1-\frac{\rho}{\rho_{c}'}\right)}
\end{equation}
Using equation (\ref{cyclic5.13}) and (\ref{cyclic5.33}) we get,
\begin{equation}\label{cyclic5.34}
\frac{dM}{M^{2}}=\frac{D'}{\left(\sqrt{\frac{\rho}{3}+\frac{1}{4r_{c}^{2}}}+\frac{\epsilon}{2r_{c}}\right)\sqrt{1-\frac{\rho}{\rho_{c}'}}}
\end{equation}
Integrating the above equation we get,
$$\frac{1}{M}=-\frac{1}{M_{i}}-D'\left[\sqrt{3\rho_{c}'}\left(\arctan\left(\frac{3A-\rho_{c}'+2\rho}{2\sqrt{\left(\rho_{c}'-\rho\right)\left(3A-\rho\right)}}\right)-\arctan\left(\frac{3A-\rho_{c}'+2\rho_{i}}{2\sqrt{\left(\rho_{c}'-\rho_{i}\right)\left(3A-\rho_{i}\right)}}\right)\right)\right.$$
$$\left.+\frac{3B\sqrt{\rho_{c}'}}{\sqrt{3B^{2}-3A-\rho_{c}'}}\left(\arctan\left(\frac{9A^{2}+3B^{2}\left(\rho_{c}'-2\rho\right)+\rho\rho_{c}'+3A\left(\rho+\rho_{c}'-3B^{2}\right)}{2B\sqrt{3\rho_{c}'\left(3A+\rho\right)\left(3B^{2}-3A-\rho_{c}'\right)\left(1-\frac{\rho}{\rho_{c}'}\right)}}\right)\right.\right.$$
$$\left.\left.-\arctan\left(\frac{9A^{2}+3B^{2}\left(\rho_{c}'-2\rho_{i}\right)+\rho_{i}\rho_{c}'+3A\left(\rho_{i}+\rho_{c}'-3B^{2}\right)}{2B\sqrt{3\rho_{c}'\left(3A+\rho_{i}\right)\left(3B^{2}-3A-\rho_{c}'\right)\left(1-\frac{\rho_{i}}{\rho_{c}'}\right)}}\right)\right)\right.$$
\begin{equation}\label{cyclic5.35}
\left.+\frac{6B\sqrt{\rho_{c}'}}{\sqrt{3A-3B^{2}+\rho_{c}'}}\left(\arctan\left(\frac{\sqrt{\rho_{c}'\left(1-\frac{\rho}{\rho_{c}'}\right)}}{\sqrt{3A-3B^{2}+\rho_{c}'}}\right)-\arctan\left(\frac{\sqrt{\rho_{c}'\left(1-\frac{\rho_{i}}{\rho_{c}'}\right)}}{\sqrt{3A-3B^{2}+\rho_{c}'}}\right)\right)\right]
\end{equation}
Now as in the previous section, $\rho_{i}\ll\rho_{c}'$, which
implies that $\frac{\rho_{i}}{\rho_{c}'}\rightarrow 0$. Hence from
the above equation we obtain,

$$\frac{1}{M}\simeq-\frac{1}{M_{i}}-D'\left[\sqrt{3\rho_{c}'}\left(\arctan\left(\frac{3A-\rho_{c}'+2\rho}{2\sqrt{\left(\rho_{c}'-\rho\right)\left(3A-\rho\right)}}\right)-\arctan\left(\frac{3A-\rho_{c}'}{2\sqrt{3A\rho_{c}'}}\right)\right)\right.$$
$$\left.+\frac{3B\sqrt{\rho_{c}'}}{\sqrt{3B^{2}-3A-\rho_{c}'}}\left(\arctan\left(\frac{9A^{2}+3B^{2}\left(\rho_{c}'-2\rho\right)+\rho\rho_{c}'+3A\left(\rho+\rho_{c}'-3B^{2}\right)}{2B\sqrt{3\rho_{c}'\left(3A+\rho\right)\left(3B^{2}-3A-\rho_{c}'\right)\left(1-\frac{\rho}{\rho_{c}'}\right)}}\right)\right.\right.$$
$$\left.\left.-\arctan\left(\frac{9A^{2}+3B^{2}\rho_{c}'+3A\left(\rho_{c}'-3B^{2}\right)}{2B\sqrt{9A\rho_{c}'\left(3B^{2}-3A-\rho_{c}'\right)}}\right)\right)\right.$$
\begin{equation}\label{cyclic5.36}
\left.+\frac{6B\sqrt{\rho_{c}'}}{\sqrt{3A-3B^{2}+\rho_{c}'}}\left(\arctan\left(\frac{\sqrt{\rho_{c}'\left(1-\frac{\rho}{\rho_{c}'}\right)}}{\sqrt{3A-3B^{2}+\rho_{c}'}}\right)-\arctan\left(\frac{\sqrt{\rho_{c}'}}{\sqrt{3A-3B^{2}+\rho_{c}'}}\right)\right)\right]
\end{equation}
At the turnaround $\rho=\rho_{c}'$. Hence we get,
$$\frac{1}{M_{c}}\simeq-\frac{1}{M_{i}}-D'\left[\frac{\pi}{2}\sqrt{3\rho_{c}'}-\arctan\left(\frac{3A-\rho_{c}'}{2\sqrt{3A\rho_{c}'}}\right)\right.$$
$$\left.+\frac{3B\sqrt{\rho_{c}'}}{\sqrt{3B^{2}-3A-\rho_{c}'}}\left(\frac{\pi}{2}-\arctan\left(\frac{9A^{2}+3B^{2}\rho_{c}'+3A\left(\rho_{c}'-3B^{2}\right)}{2B\sqrt{9A\rho_{c}'\left(3B^{2}-3A-\rho_{c}'\right)}}\right)\right)\right.$$
\begin{equation}\label{cyclic5.37}
\left.-\frac{6B\sqrt{\rho_{c}'}}{\sqrt{3A-3B^{2}+\rho_{c}'}}\arctan\left(\frac{\sqrt{\rho_{c}'}}{\sqrt{3A-3B^{2}+\rho_{c}'}}\right)\right]
\end{equation}
As before the above equation gives the remanent mass of the BH at
the turnaround. For $M_{i}\gg M_{p}$, $M_{c}$ becomes independent
of $M_{i}$ and is given by the following relation,
$$\frac{1}{M_{c}}\simeq D'\left[\arctan\left(\frac{3A-\rho_{c}'}{2\sqrt{3A\rho_{c}'}}\right)\right.$$
$$\left.-\frac{3B\sqrt{\rho_{c}'}}{\sqrt{3B^{2}-3A-\rho_{c}'}}\left(\frac{\pi}{2}-\arctan\left(\frac{9A^{2}+3B^{2}\rho_{c}'+3A\left(\rho_{c}'-3B^{2}\right)}{2B\sqrt{9A\rho_{c}'\left(3B^{2}-3A-\rho_{c}'\right)}}\right)\right)\right.$$
\begin{equation}\label{cyclic5.38}
\left.+\frac{6B\sqrt{\rho_{c}'}}{\sqrt{3A-3B^{2}+\rho_{c}'}}\arctan\left(\frac{\sqrt{\rho_{c}'}}{\sqrt{3A-3B^{2}+\rho_{c}'}}\right)-\frac{\pi}{2}\sqrt{3\rho_{c}'}\right]
\end{equation}

\section{Hawking evaporation of black holes}
From the above evaluation it is clear that there is always a
remanent mass of the BHs at the turnaround in an oscillating
universe. So contrary to existing literature we conclude that
there is no possibility of destruction of BH . The BHs formed
during the expanding cycle of the cyclic universe survive into the
next cycle and eventually grow in size. Therefore they create
undesired non-uniformity in a nearly uniform universe. Eventually
the BHs will occupy the entire volume of the horizon and will be
responsible for the destruction of the cyclic models. This is a
serious defect indeed! Hence the problem posed by the BHs in a
cyclic universe still stands un-eliminated.

But in \cite{Brown1} it has been argued that for a BH with mass
$M=10^{5}M_{p}$, Hawking evaporation takes place in time
$\tau\sim\frac{25\pi M^{3}}{M_{p}^{4}}\sim10^{-27}$ sec and
ultimately the BH becomes non-existent, thus causing no problems.
Here $M_{p}$ represents Planck mass. In our above calculations we
have considered $G=M_{p}^{-2}=1$. Rewriting equation
(\ref{cyclic5.18}), for RSII brane we get,
\begin{equation}\label{cyclic5.39}
M_{c}\simeq\frac{M_{i}}{1+\frac{DM_{i}\sqrt{2\lambda}\log{\left(\frac{\lambda+\rho_{c}+\sqrt{\alpha+2\lambda\rho_{c}+\rho_{c}^{2}}}{\lambda}\right)}}{M_{p}^{3}}}
\end{equation}
$D$ is taken as a constant of the order unity and
$\log{\left(\frac{\lambda+\rho_{c}+\sqrt{\alpha+2\lambda\rho_{c}+\rho_{c}^{2}}}{\lambda}\right)}\sim
M_{p}^{2}$. Then we have
$M_{c}\sim\frac{M_{i}}{1+\frac{M_{i}}{M_{p}}}$. For a BH with
$M_{i}\gg M_{p}$, $M_{c}\sim M_{p}$, i.e., the remanent mass of
the BH is of the order of Planck mass. Hence the remanent BH
undergoes Hawking evaporation in time $\tau\sim10^{-43}$ sec at
the order of Planck time. A similar evaluation is possible for DGP
brane model as well. So fortunately, the remanent BHs do not cause
any problems. This gives a possible solution of the BH problem in
cyclic universe described by brane-world scenario.

\section{Conclusion and Discussion}
A serious problem is posed by the existence of black holes in an
oscillating universe. In \cite{Brown1} Brown et al suggested that
in an oscillating cosmology the black holes keep losing mass due
to phantom energy accretion before totally disappearing before the
turnaround. Babichev et al in \cite{Babichev1} devised a
successful mechanism which was in accordance with the result given
by Brown et al. In this paper we have investigated the outcome of
phantom energy accretion on black holes in a cyclic universe
described by brane-worlds. It is seen that RSII brane model
readily incorporates the oscillating nature in its framework. The
only condition being the negativity of the brane tension,
$\lambda$ or the bulk cosmological constant, $\Lambda_{4}$. But
unlike RSII brane model, DGP brane does not readily support the
cyclic nature of the universe. So a modified DGP brane equation,
that supports cyclic turnarounds has been proposed for our
evaluation.

It is found that for RSII brane model, during the expanding phase
the black hole mass gradually decrease with the increase of
phantom energy density and finally reaches a critical value at
which the turnaround occurs. This result is absolutely consistent
with the known results in literature. In the contracting phase the
black hole mass again decreases with the increase in the
non-phantom components of the universe. This is however contrary
to our expectations. So it is understood that our evaluations
after the turnaround are not really rigid. We see that the black
holes in a cyclic universe reaches a remanent mass, $M_{c}$ before
turnaround. So, the remanent mass implies that the destruction of
black holes is not a real possibility in the cyclic cosmology,
with phantom energy turnarounds, for a universe characterized by
brane gravity. However fortunately we find that the remanent
masses of black holes at turnaround do not cause problems. The
reason being that these remanent black holes Hawking evaporate in
a time $\tau\sim10^{-43}$.

\vspace{6mm}
{\bf Acknowledgement:}\\\\ The author acknowledges Mr.Ritabrata
Biswas and Dr.Ujjal Debnath for helpful discussions. The author
also thanks the anonymous referee for useful comments on the
manuscript.

\end{document}